\documentclass{aa}
\usepackage{graphicx}

\begin{document}

   \title{On the light echo in V838 Mon}


   \author{R. Tylenda}

   \offprints{R. Tylenda, email: tylenda@ncac.torun.pl}

   \institute{N. Copernicus Astronomical Center, Department for Astrophysics,
                 Rabia\'nska 8, 87--100 Toru\'n, Poland}

   \date{Received ; accepted }

   \titlerunning{Light echo in V838 Mon}
   \authorrunning{R. Tylenda}

\abstract{After having presented a theoretical outline of the light echo
phenomenon and
results of simple numerical simulations we study the available images of
the light echo in V838~Mon obtained with HST. From an analysis of the observed 
expansion of the light echo we conclude that the distance to V838~Mon is greater
then $\sim$5~kpc. We also investigate the structure of the dust distribution
in the vicinity of the object. We find no obvious signs of spherical symmetry
in the resultant distribution. Near the central object there is a strongly
asymmetric dust-free region which we interpret as an evidence that V838~Mon
is moving relatively to the dusty medium. From these results we conclude
that dust illuminated by the light echo is of interstellar origin rather
than produced by mass loss from V838~Mon in the past.

}

   \maketitle

\section{Introduction}

\object{V838 Mon} was discovered in eruption in the beginning of 
January 2002 (Brown \cite{brown}). 
The main outburst, however, started in the beginning of
February 2002 and, in the optical, lasted for about two months (e.g. Munari
et~al. \cite{munari}). The nature of the eruption is enigmatic (e.g. Munari
et~al. \cite{munari}, Kimeswenger et~al. \cite{kimes}). It cannot be
accounted for by known thermonuclear events (classical nova, late He-shell
flash) so new, unexplored so far, mechanisms have been searched for (Soker \&
Tylenda \cite{soktyl}, Retter \& Marom \cite{retmar}).

\object{V838 Mon} has recently received significant publicity, even in non-scientific
media, due to its light echo. This event has been discovered shortly
after the main eruption in February 2002 (Henden et~al. \cite{henden}).
However, the most spectacular images of the \object{V838~Mon} light echo have been
provided in later epochs by HST (Bond et~al. \cite{bond}).

The light echo is a rare event and can be observed when a light outburst
illuminates circumstellar or interstellar dust. So far it has primarily been 
observed for extragalactic supernovae 
(e.g. Chevalier \cite{cheval}, Xu et al. \cite{xu}). In our Galaxy this
phenomenon has been observed in \object{Nova Persei 1901} 
(e.g. Couderc \cite{couderc}).

Analyses of light echoes can be used while investigating supernova light curves
(e.g. Chevalier \cite{cheval}). Observed evolution of the light echo arround
\object{SN 1987A} in LMC has enabled to study the dust distribution in front of the
supernova up to a distance of $\sim$1~kpc (see e.g. Xu et al. \cite{xu}).
Supernova light echoes can also be used to measure distances to galaxies
(Sparks \cite{sparks}).
 
Munari et~al. (\cite{munari}) and Kimeswenger et~al. (\cite{kimes})
have attempted to use the observed expansion of the light echo arround 
\object{V838~Mon} to measure the distance. Their calulations have however been based
on a naive interpretation of the light echo expansion (i.e. that it
expands at the velocity of light) and the derived values 
of 0.6--0.8~kpc are in fact significant underestimates of the distance. A more
realistic analysis in Bond et~al. (\cite{bond}) gave a lower limit to the
distance of $\sim$6~kpc.

In this paper we present an analysis of the available data on the light echo
in \object{V838~Mon}. After presenting theoretical considerations 
and results of
simple simulations of light echoes (Sect.~2) we attempt to use the
observational data to constrain the distance to \object{V838~Mon} (Sect.~3) and to study the
dust distribution in front of the object (Sect.~4).

\section{Basic considerations and numerical simulations  \label{basic}}

Couderc (\cite{couderc}) was probably the first to give a correct explanation for the
light echo arcs observed arround \object{Nova Persei 1901}. More recently, 
theoretical considerations on the light echo can be found in
papers devoted to supernova light echoes (see e.g. Chevalier
\cite{cheval}). The subject is however relatively unknown in the field of 
stellar astrophysics. This is probably the reason why
in some papers oversimplified or even incorrect
interpretations of the \object{V838 Mon} light echo can be found. Therefore
let us summarize 
basic formulae describing the structure and evolution of the light echo.

Let us define a rectangular ($x,y,z$) coordinate system with 
its origin being in the light source. The $z$ axis is along the line of sight 
toward the observer. The ($x,y$) plane is perpendicular to the $z$ axis, 
i.e. is tangent to the sky sphere in the source. Let the source emits a
short light flash at a time $t = 0$. We assume the so-called single
scattering approximation, i.e. that photons are scattered only once in the
dusty medium surrounding the source.
Suppose that the scattered echo is seen 
by the observer at a time $t$. Then the illuminated dust lies on a surface
defined by 
\begin{equation}
  r + l = d + ct  \label{elipse}
\end{equation}
where $c$ is the velocity of light, $d$ is the distance between the source and
the observer, $r = \sqrt{ x^2 + y^2 + z^2 }$ is the distance of the
scaterring dust from the source, and $l = \sqrt{ x^2 + y^2 + (d - z)^2}$
is the distance of the dust from the observer. Eq.~(\ref{elipse}) is an
ellipsoid having foci in the source and the observer.

The considerations can be simplified when $x$ and $y$ are much smaller than
$d$, i.e. when the observed angular dimensions of the echo are small. 
Putting $l = d - z$ Eq.~(\ref{elipse}) becomes
\begin{equation}
  x^2 + y^2 = (ct)^2 + 2zct.
  \label{parab_e}
\end{equation}
This is a paraboloid symmetric arround the $z$ axis (line of sight) and
having focus in the source. In the
case of \object{V838~Mon} the echo has dimensions of order of $1\arcmin$ so
in this case the accuracy of Eq.~(\ref{parab_e}) is better than $10^{-3}$.

\begin{figure}
  \resizebox{\hsize}{!}{\includegraphics{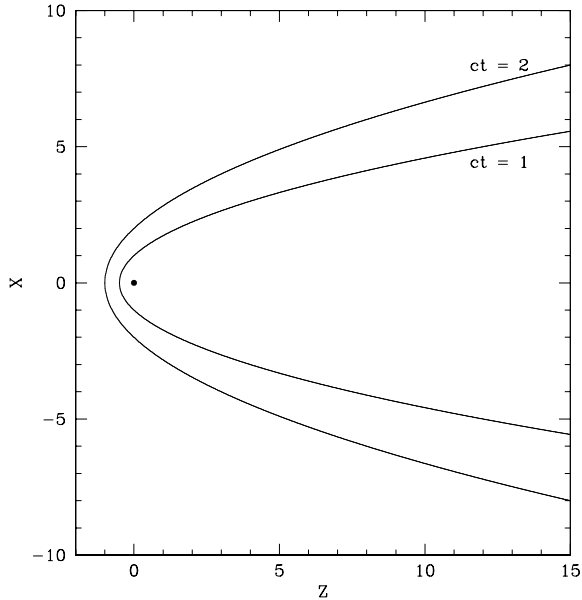}}
  \caption{The light echo paraboloid (Eq. \ref{parab_e}) shown in the $xz$ plane 
        for $ct = 1$ and $ct = 2$. The axes are in units of $ct$. The dot marks the
        source position at $x = 0, z = 0$. The observer is at $x = 0, z = \infty$.}
  \label{parab_f}
\end{figure}

Figure \ref{parab_f} shows Eq. (\ref{parab_e}) in the $xz$ plane for $ct = 1$
and $ct = 2$ (axes in the figure are in units of $ct$). 
Several important conclusions can be drawn from Fig.~\ref{parab_f}.
If the source (at $x = 0, z = 0$) has produced a short flash of light at $t = 0$
then at $t = 1/c$ and $t = 2/c$ the observer would see dust illuminated on
paraboloids $ct = 1$ and $ct = 2$, respectively. However, if the flash
started at $t = 0$ and lasted up to $t = 1/c$ then at $t = 1/c$ all dust 
within the paraboloid $ct = 1$ would be illuminated for the observer. 
At $t = 2/c$ illuminated dust would be between the paraboloids $ct =1$ and
$ct = 2$. Thus
contrary to some naive considerations the echo does not expand
spherically starting from the source. It starts from the line of sight, i.e.
$x = 0, z \ge 0$, and most effectively penetrates regions situated 
in front of the source. Behind the source the penetration is slowest.

The observed echo is in the ($x,y$) plane and its structure and evolution 
depends on the dust distribution arround the source. 
One of important conclusions which can
easily be drawn from Fig.~\ref{parab_f} is that
if the observed echo has a well defined outer rim (as observed in the case of
\object{V838~Mon}) the dusty medium arround the source has to have a rather well 
defined outer boundary in front of the source. 

In order to illustrate
dependence of the light echo structure and evolution on the dust
distribution let us consider two simple dust geometries, i.e. a plane
paralel slab and a spherical shell centred on the source.

\subsection{Plane paralel dust slab  \label{slabs}}

Let us assume that a thin plane paralel slab of dust intersects the line of sight 
at $z_0$ and that the normal to the slab is inclined to the $z$ axis 
at an angle $\alpha$. Let us also assume, for simplicity, that the normal lies in
the ($x,z$) plane. The slab is then described by
\begin{equation}
  z = z_0 - ax   \label{slab}
\end{equation}
where $a = \tan \alpha$. Eq.~(\ref{slab}), when used to eliminate $z$ from
Eq.~(\ref{parab_e}), gives the echo shape as follows
\begin{equation}
  x^2 + y^2 = (ct)^2 + 2z_0ct - 2axct.  \label{echo_slab}
\end{equation}
Thus the echo is in the shape of a ring whose radius, $r_\mathrm{e}$, is
given by
\begin{equation}
  r_\mathrm{e} = \sqrt{(1 + a^2)(ct)^2 + 2z_0ct}.  \label{r_echo_slab}
\end{equation}
The ring is centred at
\begin{equation}
  x_\mathrm{c} = -act, \, y_\mathrm{c} = 0.  \label{c_echo_slab}
\end{equation}
The velocity of expansion of the echo ring, $v_\mathrm{e}$, is
\begin{equation}
  v_\mathrm{e} \equiv \frac{\mathrm{d} r_\mathrm{e}}{\mathrm{d} t} = 
    \frac{(1 + a^2)c^2t + z_0c}{r_\mathrm{e}} =
    \frac{(1 + a^2)c^2t + z_0c}{\sqrt{(1 + a^2)(ct)^2 + 2z_0ct}}.
  \label{v_echo_slab}
\end{equation}

An analysis of
Eqs. ~(\ref{r_echo_slab} -- \ref{v_echo_slab}) leads to the following
conclusions.
The echo ring is centred on the
source only if the dust slab is perpendicular to the line of sight.
In almost all the cases the echo expansion is superluminal.
Only if $z_0 = 0$ and $a = 0$ the echo expands at $c$.
For $z_0 > 0$ (the slab in front of the source) the echo appears at the same
time as the
source flash, i.e. at $t = 0$. It starts expanding from the source position, 
i.e. from $x_\mathrm{c} = 0, y_\mathrm{c} = 0$. 
For $z_0 < 0$ (the slab behind the source)
the echo appears with a time delay relative to the source flash, 
i.e. it starts at $t = (-2z_0)/((1+a^2)c)$. In this case it begins expanding
from the position $x_\mathrm{c} = (2z_0a)/(1+a^2), y_\mathrm{c} = 0$.
In both cases the expansion velocity is $\infty$ at the moment of the echo
appearance, i.e. at $t = 0$ if $z_0 > 0$ and at $t = (-2z_0)/((1+a^2)c)$
if $z_0 < 0$. It decreases asymptotically to $\sqrt{1 + a^2}c$ when 
$(1 + a^2)ct \gg |z_0|$. For slabs non-perpendicular to the line of sight,
i.e. when $a \neq 0$, the echo centre moves away from the source at a constant
velocity of $|a|c$. The direction of this movement is opposite to the
direction of the slab normal projected on the sky surface.

\subsection{Spherical dust shell  \label{shells}}

Let us assume that a thin spherically symmetric shell of dust having 
a radius, $r_0$, is centred on the source, so it is described by
\begin{equation}
  r_0^2 = x^2 + y^2 + z^2.   \label{shell}
\end{equation}
This when used to eliminate $z$ from Eq.~(\ref{parab_e}) gives
\begin{equation}
   x^2 + y^2 = 2r_0ct - (ct)^2.  \label{echo_shell}
\end{equation}
Thus the echo is in the shape of a ring with the radius, $r_\mathrm{e}$,
given by
\begin{equation}
   r_\mathrm{e} = \sqrt{2r_0ct - (ct)^2}  \label{r_echo_shell}
\end{equation}
and centred on the source ($x_\mathrm{c} = 0, y_\mathrm{c} = 0$). 
The echo expansion velocity, $v_\mathrm{e}$, is
\begin{equation}
  v_\mathrm{e} \equiv \frac{\mathrm{d} r_\mathrm{e}}{\mathrm{d} t} = 
     \frac{r_0c - c^2t}{r_\mathrm{e}} =
     \frac{r_0c - c^2t}{\sqrt{2r_0ct - (ct)^2}}.
  \label{v_echo_shell}
\end{equation}

As can be seen from Eq.~(\ref{r_echo_shell}), the echo radius
starts from zero at $t = 0$, reaches a maximum equal to $r_0$
at $t = r_0/c$, and collapses to zero at $t = 2r_0/c$. 
Note that in a coordinate system ($ct-r_0, r_\mathrm{e}$) 
Eq.~(\ref{r_echo_shell}) describes a circle centred on the origin and having
the radius equal to $r_0$. 
The expansion velocity of the echo ring starts from $\infty$ at 
$t = 0$, decreases and reaches zero at $t = r_0/c$, and continues decreasing to
$-\infty$ at $t = 2r_0/c$.

\subsection{Numerical simulations}

In order to illustrate the evolution of the echo in different geometries of
the dust distribution we have performed simple numerical simulations based
on the theoretical considerations outlined above. The
structure of the light echo has been determined in the paraboloid
approximation, i.e. using Eq.~(\ref{parab_e}). If the
central source radiates at a luminosity, $L_\nu$, the intensity, $I_\nu$, of
the radiation scattered at a point ($x,y,z$) after the time delay, $t = r/c$
(where $r = \sqrt{x^2 + y^2 + z^2}$), can be calculated from
\begin{equation}
   I_\nu = \frac{\sigma_\nu L_\nu}{16\pi^2r^2}.  \label{intens}
\end{equation}
Here $\sigma_\nu$ is the scattering coefficient. Eq.~(\ref{intens})
adopts the single scattering approximation and that the scattering is isotropic. 
Integrating Eq.~(\ref{intens}) 
along the $z$ axis over the dusty medium illuminated at a given time, $t$,
gives the echo surface brightness at the point ($x,y$) at this time moment.

\begin{figure*}
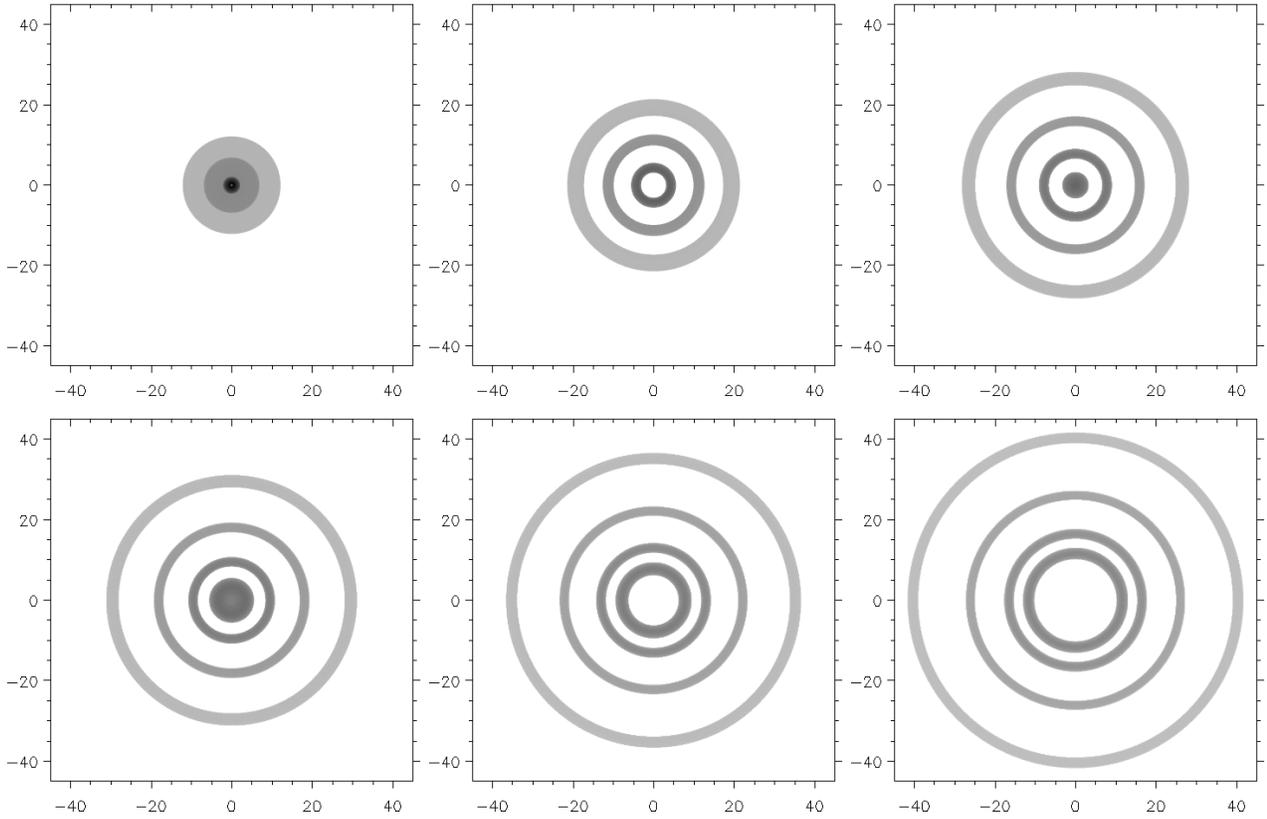

\centering
  \includegraphics[width=5.5cm]{sheets50.eps2}
  \includegraphics[width=5.5cm]{sheets150.eps2}
  \includegraphics[width=5.5cm]{sheets250.eps2}
  \includegraphics[width=5.5cm]{sheets300.eps2}
  \includegraphics[width=5.5cm]{sheets400.eps2}
  \includegraphics[width=5.5cm]{sheets500.eps2}
  \caption{The evolution of the echo structure produced by four dust slabs
           perpendicular to the line of sight at the $z$ positions of 
           $-0.10$, 0.00, 0.30, and 1.00~pc. The source (not shown in the
           figure but lying in the centre of each image) has experienced a
           50~days long flash of constant luminosity. All the slabs have the same
           thickness of 0.02~pc and the same value (constant within the
           slabs) of the scattering coefficient $\sigma_\nu$. The object is
           assumed to be observed from a distance of 5~kpc. The images are
           ordered from upper--left to lower--right and correspond to the time
           epochs of 50, 150, 250, 300, 400, and 500 days after the beginning 
           of the source flash. The axes of the images are in arcsec. 
           The grey scale is logarithmic and is normalized to the
           brightest region in the figure.
           The brightest centre of the first image and the faintest 
           outer ring in the last images span $\sim$4 orders of magnitude in 
           the surface brightness.}
  \label{sheets}
\end{figure*}

The simulations have been done for two simple
geometries of the dust distribution, similarily as above. 
In the case presented in
Fig.~\ref{sheets} dust has been adopted to be uniformly (i.e. having 
constant $\sigma_\nu$) distributed in four
plane paralel slabs perpendicular to the $z$ axis. The $z_0$ positions of the
slabs are $-0.10$, 0.00, 0.30, and 1.00~pc. The slab thickness is the
same in all cases and is equal to 0.02~pc. $L_\nu$ has been adopted to
be constant during the source flash which lasted for 50 days (otherwise
$L_\nu = 0$). Fig.~\ref{sheets} shows the echo structure as observed from a
distance of 5~kpc at 50, 150, 250, 300, 400, and 500 days 
after the beginning of the flash. The grey scale is logrithmic and it has
been normalized to the maximum surface brightness in the images. Between the
brightest centre of the first image and the faintest outer ring in the last
image there is a span of 4 orders of magnitude in the surface brightness.

The first image in Fig.~\ref{sheets} shows the echo just at the end of the
source flash. According to the discussion of Fig.~\ref{parab_f} all dust
being within the paraboloid defined by the time elapsed since the beginning
of the flash, is illuminated. Thus the echo is composed of three filled
circles produced by three slabs having $z_0 \ge 0.0$ (the fourth slab at 
$z_0 = -0.1$~pc is not seen as it has not yet been reached by the light
paraboloid). Later on the illuminated dust is situated between 
two paraboloids,
i.e. the first one corresponding to the begining of the flash 
and the second one defined by the epoch of the end of the flash.
Thus in the second image, taken 100~days after the end of the flash, the echo
consists of three rings. The slab at $z_0 = -0.10$~pc
is reached by the first paraboloid at $t \simeq 210$~days, whereas the
second paraboloid starts leaving it at $t \simeq 310$~days. Therefore this
slab is seen in the form of a central disc in the third and fourth image.
Later on it produces the innermost ring in the image. 

As can be
seen from Fig.~\ref{sheets} the surface brightness of the rings decreases
with time. For geometrical reasons the decrease is strongest for the slab at
$z_0 = 0.0$. From the first image to the last one in Fig.~\ref{sheets} the
surface brightness in the ring corresponding to $z_0 = 0.0$ decreases by
factor $\sim$500 whereas the same drop in the case of the outermost ring 
($z_0 = 1.00$~pc) is only factor $\sim$2. With time the echo rings would
continue expanding with slowly decreasing surface brightness and with 
the expansion velocity approaching $c$. 

\begin{figure*}
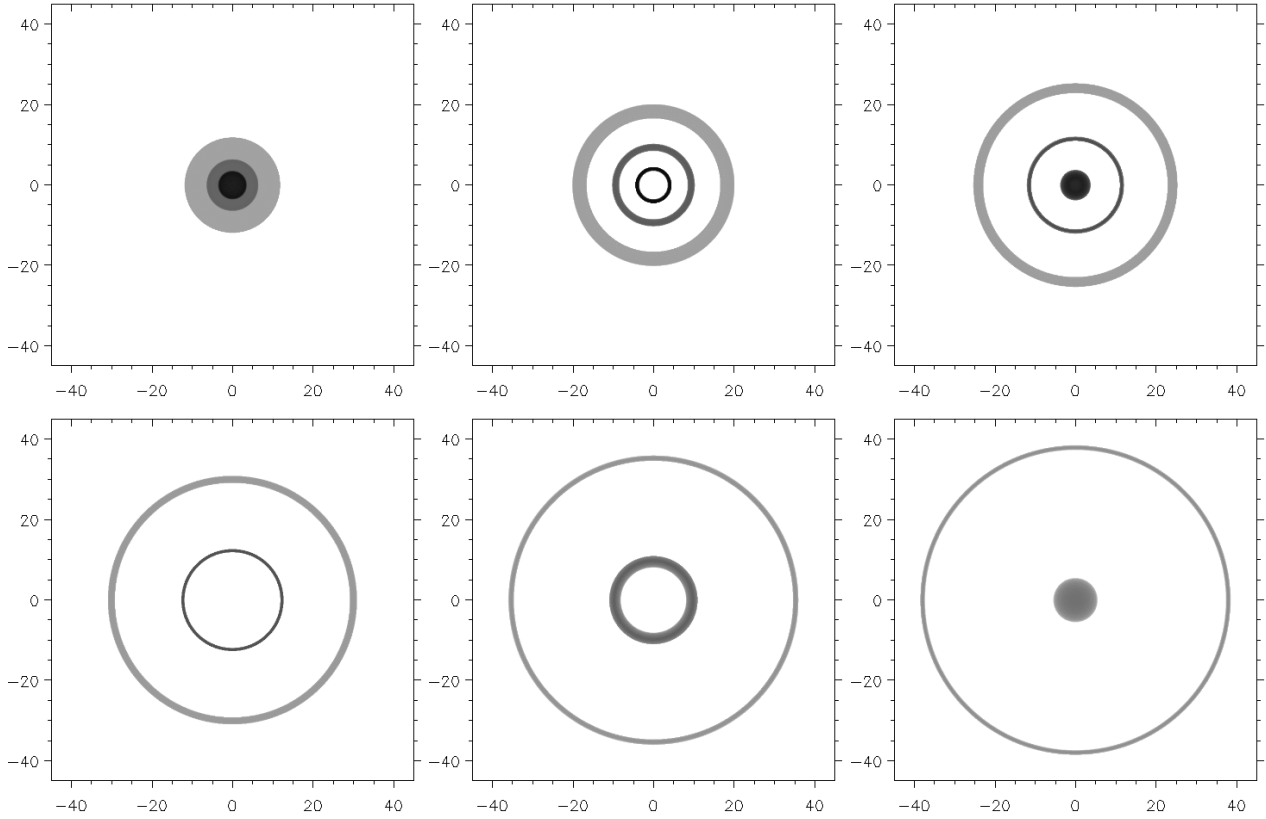

\centering
  \includegraphics[width=5.5cm]{spheres50.eps2}
  \includegraphics[width=5.5cm]{spheres150.eps2}
  \includegraphics[width=5.5cm]{spheres250.eps2}
  \includegraphics[width=5.5cm]{spheres400.eps2}
  \includegraphics[width=5.5cm]{spheres600.eps2}
  \includegraphics[width=5.5cm]{spheres750.eps2}
  \caption{The same as Fig.~\ref{sheets} but for three spherical shell of
           dust centred on the source. The shells have the same thickness 
           of 0.02 pc and radii of 0.10, 0.30, and 1.00~pc. The images shows 
           the echo as observed 50, 150, 250, 400, 600, and
           750~days after the beginning of the source flash. 
           The brightest and the faintest regions in the figure span 
           $\sim$2.5 orders of magnitude in the surface brightness.}
  \label{spheres}
\end{figure*}

Figure~\ref{spheres} shows the evolution of the echo structure produces by
three spherically symmetric dust shells centred on the source. The radii of
the shells are 0.10, 0.30, and 1.00~pc. All the shells have the same
thickness equal to 0.02~pc. As in the previous case the source flash has lasted
for 50~days and the object is observed from a distance of 5~kpc. The
consecutive images correspond to the epochs of 50, 150, 250, 400, 600, and
750~days after the beginning of the source flash.

As in Fig.~\ref{sheets} the first image in Fig.~\ref{spheres} corresponds to
the end of the source flash and all dust within the light paraboloid
produced by the beginning of the flash is illuminated. The echo is thus
composed of three filled circles produced by illuminated sections of the
corresponding shells. The echo components initially expand superluminally.
However, as discussed in Sect.~(\ref{shells}),
the echo from a shell slows the
expansion rate down to 0 when reaching the maximum radius equal to 
the value of $r_0$. For
the shells in our simulations, i.e. having $r_0 = 0.1$, 0.3, and 1.0~pc, 
it happens at $t \simeq 130$, 370, and 1200~days, respectively. At the same
time the echo ring gets its maximum surface brightness and minimum
thickness. Next the echo ring starts shrinking and finally disappears as a
fast collapsing disc. In our simulations the echoes due to the shells
having $r_0 = 0.1$, 0.3, and 1.0~pc disappear at $t = 310$, 790, and
2450~days. In the second image in Fig.~\ref{spheres} the innermost ring is
close to its maximum radius whereas in the third image the echo from the
innermost shell is collapsing. The echo form the shell having $r_0 = 0.3$~pc
is close to its maximum radius in the fourth image whereas its collapse is
observed in the last two images.

\subsection{Discussion}

The above considerations, although done for simlified cases,
demonstrate that long term observations of the evolution of the echo
structure can provide information on the dust distribution and thus,
indirectly, also on the origin of the dust producing the echo. If dust
results from a past mass loss from the central object it would tend to
show certain symmetries in its distribution around the source. Usually mass loss is
not continuous at a constant rate but rather in form of more or less
periodic fluctuations of the mass loss rate, e.g. during the AGB phase, 
or discrete events of strong mass loss separated by longer periods of
quiescence, e.g. in nova-like outbursts. Then the dust distribution would show
a succession of shell-like structures or be in form of a more or less defined
envelope. The evolution of the light echo in such circumstances would follow
general paterns discussed in Sect.~\ref{shells} and Fig.~\ref{spheres}.
The initially fast expansion, slowing down when reaching maximum sizes corresponding 
to outer edges of the dust shells or envelope would be followed by an accelerated 
collapse. 

If, however, dust is not related to the source and is rather of
interstellar character no obvious symmetry in respect to the light source is
expected to occur in its distribution. Instead dust would rather be in form
of extended sheets, zones or filaments and our considerations of plane
paralel slab
geometries in Sect.~\ref{slabs} and Fig.~\ref{sheets} would be more
appropriate. The echo would expand more and more with slowly decreasing
brightness as more and more distant regions are illuminated.

Note that for studying the general character of the dust distribution around
the source the echo has to observed for a long enough time.
The initial evolution of the echo produced by spherical dust shells is
similar to that due to slabs. The light paraboloid is close to the line
of sight so the echo does not feel differences in the general shape of the
dust distribution. This can be seen from Eqs.~(\ref{r_echo_slab}) and
(\ref{r_echo_shell}) which reduce to the same form if $ct \ll z_0$ and
$ct \ll r_0$, respectively.
Thus the first image in Fig.~\ref{spheres} is
not much different from that in Fig.~\ref{sheets} (the reason why the
innermost (brightest) disc in Fig.~\ref{sheets} is significantly smaller
than the corresponding one in  Fig.~\ref{spheres} is that in the former case it
is produced by the slab at $z_0 = 0.0$ whereas in the latter case the shell
has $r_0 = 0.1$~pc). Later on, when $ct$ is no more negligible compared to
$z_0$ and $r_0$ the differences between the geometries are
better and better seen in the echo evolution.

\section{Distance to V838 Mon constrained from the echo expansion \label{dist}}

An analysis of the light echo evolution can be used to estimate the
distance to the object. One possible way of doing that 
using the observed polarization structure of the echo has been proposed in
Sparks (\cite{sparks}). An attempt to derive the distance to \object{V838~Mon} has
been done in Bond et~al. (\cite{bond}). They measured expansion rates of
ring-like structures in the echo from two consecutive images and found
that the distance is greater than 2~kpc. An attempt to use the method
proposed by Sparks (\cite{sparks}) led Bond et~al. to conclude that the lower
limit to the distance is $\sim$6~kpc.

We have attempted to put constrains on the distance to \object{V838~Mon} from
measuring expansion of the outer edge of the echo. As it is clear from the
considerations in Sect.~\ref{basic} the outer edge of the echo is produced
by the beginning of the light flash reflected at the outermost edge of the dust
distribution in front of the source. 

For the plane slab geometry
Eq.~(\ref{r_echo_slab}) when combined with Eq.~(\ref{c_echo_slab}) can be rewritten as
\begin{equation}
  \theta = \frac{r_\mathrm{e}}{d} = 
           \sqrt{(\frac{2z_0}{ct} + 1)(\frac{ct}{d})^2 + \theta_\mathrm{c}^2}
  \label{teta_echo_slab}
\end{equation}
where $d$ is the distance to
the object, $\theta$ is the angular radius of the echo, and 
$\theta_\mathrm{c} = x_\mathrm{c}/d$ is the
angular distance of the echo centre from the source. 
At early epochs, i.e. when $(2z_0)/(ct) \gg 1$,
Eq.~(\ref{teta_echo_slab}) allows to determine only the value of $z_0/d^2$.
Thus unless $z_0$ is known the distance cannot be determined. However, in
later epochs, when $(2z_0)/(ct) \ll 1$, Eq.~(\ref{teta_echo_slab}) can be
used to derive $d$ from the observed values of $\theta$ and $\theta_\mathrm{c}$.

In the case of the spherical symmetry Eq.~(\ref{r_echo_shell}) can be rewritten as
\begin{equation}
   \theta  = \sqrt{(\frac{2r_0}{ct} - 1)(\frac{ct}{d})^2}.
   \label{teta_echo_shell}
\end{equation}
Thus, not only in early phases (fast initial expansion) but also in late
phases (final collapse) the distance cannot be determined unless the value
of $r_0$ is known. Only near the maximum size of the echo, i.e. when
$r_0 \approx ct$, Eq.~(\ref{teta_echo_shell}) can give a direct estimate of
the distance.

In a real case, such as \object{V838~Mon}, we do not know a priori what is the
distance of the outer edge of dust from the source nor
what is the geometry of this edge. Thus the only way is to look at the
observed evolution of the echo and try to fit it with Eq.~(\ref{teta_echo_slab})
or Eq.~(\ref{teta_echo_shell}).

For this purpose we have measured the positions of the outer edge of the
light echo of \object{V838~Mon} on five images taken by H.E.~Bond and available at
the HST web site (http://hubblesite.org/newscenter/archive/2003/10/,
see also Bond et~al. \cite{bond}). The images cover the period from April~30
to December~17, 2002. From each image the positions of
typically 60--70 points at the outer rim of the echo
(more or less equally spaced in the azimuthal angle) have been measured.
Note that all the maeasurements (also those reported in the next section)
have been done on the negatives of the published images as then corresponding emission
edges are more easily seen. The
positions have been determined in the coordinate system centred on \object{V838~Mon}
with the $x$ and $y$ axes pointing to west and north, respectively.
Then a circle has been fitted to the measured positions on each image using the least
square method. The results are presented in Table~\ref{echo_obs}. First column
shows the time of observations given in days since January~1, 2002. The radius of
the echo, $\theta$, and its uncertainty, $\epsilon$, 
(measured as a standard deviation of the observed points from the fitted circle) 
are given in the next two columns. The last
two columns show the position of the centre of the fitted circle relative
to the central star. All the results are in arcsec.

\begin{table}
\caption{Results of fitting a circle to the outer edge of the light echo of
         V838 Mon. Time is in days since 1 January 2002. Results are in arcsec.}
\label{echo_obs}
\begin{tabular}{c c c c c}
\hline
\multicolumn{1}{c}{$t$(days)} &
\multicolumn{1}{c}{$\theta$} &
\multicolumn{1}{c}{$\epsilon$} &
\multicolumn{1}{c}{$x$(centre)} &
\multicolumn{1}{c}{$y$(centre)} \\
\hline
 120.0 & 18.55 & 1.08 & -0.84 & 0.46 \\
 140.0 & 20.89 & 1.05 & -1.12 & 0.51 \\
 245.0 & 30.32 & 1.44 & -2.27 & 0.37 \\
 301.0 & 33.68 & 1.46 & -2.36 & 1.11 \\
 351.0 & 36.66 & 1.56 & -2.61 & 1.58 \\
\hline
\end{tabular}
\end{table}

In order to make a quantitative analysis of the results in Table~\ref{echo_obs} 
it is neccesary to determine the time moment of the zero age of the echo,
$t_0$.
The echo has been discovered in mid-February 2002 (Henden et~al.
\cite{henden}) and it has been suggested that the observed echo results from 
the main outburst which started in the beginning of February 2002 
(Munari et~al. \cite{munari}, Bond et~al. \cite{bond}).
This is supported by the fact that in the HST images the outer echo rim is blue and
it was in the beginning of the main outburst when the star was bluest (see
e.g. Bond et~al. \cite{bond}). The main outburst started on February~1 and
the maximum has been reached on February~5--6 (Munari et~al. \cite{munari}, 
Bond et~al. \cite{bond}). Thus we adopt February~3 as the date of the zero age 
of the echo, i.e. $t_0 = 34$~days.

As can be seen from Table~\ref{echo_obs}, the centre of the echo migrates
from the central star. The migration keeps more or less the same direction 
and the distance between the echo centre and the star
increases roughly linearly with time since the zero age.
This is what is expected from the plane geometry of the dust
distribution (see Eq.~\ref{c_echo_slab}). It cannot be reconciled with a
spherically symmetric distribution. Thus
the shape of the outer edge of dust in front of the star can be approximated
by a plane inclined to the line of sight rather then a sphere centred on the
star.

A linear fit to the observed evolution of the distance of the echo centre 
from the star as given in Table~\ref{echo_obs} results in a relation
\begin{equation}
  \label{cent_obs}
  \theta_\mathrm{c} = (0\farcs0106 \pm 0\farcs0008)(\frac{t}{1\,\mbox{day}})
\end{equation}
where $t$ is the time since the zero age of the echo. This with
Eq.~(\ref{c_echo_slab}) allows to derive the inclination of the dust surface
to the line of sight, i.e.
\begin{equation}
  \label{alfa}
  \tan \alpha = (0.061 \pm 0.005)(\frac{d}{1\,\mbox{kpc}}).
\end{equation}
Unfortunately the distance cannot be determined with a satisfactory accuracy
(see below). Adopting $d = 5$~kpc one obtaines $\alpha \simeq 16\degr$
whereas for $d = 10$~kpc Eq.~(\ref{alfa}) gives $\alpha \simeq 30\degr$.
Note that the azimuthal angle of the normal to the dust surface is opposite
(i.e. $+ 180\degr$) to that of the echo centre.

A $\chi^2$ minimum fit of Eq.~(\ref{teta_echo_slab}) to the observed values of
$\theta$ in Table~\ref{echo_obs} gives the best fit for $d \simeq 6.1$~kpc
and $z_0 \simeq 2.0$~pc. However the $\chi^2$ minimum is rather shallow and 
extended along a $z_0 \sim d^2$ relation. 
From a 90\% confidence level on the ($d,z_0$) plane we can only state that 
the lower limit to the distance is $\sim 2.5$~kpc.

Clearly we are well before the phase when the distance can be unambigiously 
determined, i.e. when $(2z_0)/(ct) \ll 1$. Another reason for the large
uncertainty in the distance estimate is in significant uncertainties
in the observed values of the echo radius in Table~\ref{echo_obs}. These
uncertainties are simply due to the fact that the echo rim is not ideally
reproduced by a circle. Clearly the outer edge of the dust distribution in
front of \object{V838~Mon} is not a perfect plane.

\begin{table}
\caption{Results of fitting a circle centred on the star to the north-west 
         quadrant ($x>0,\,y>0$) of the outer edge of the light echo of
         \object{V838~Mon}. Time is in days since 1 January 2002. 
         Results are in arcsec.}
\label{echo_nw}
\begin{tabular}{c c c}
\hline
\multicolumn{1}{c}{$t$(days)} &
\multicolumn{1}{c}{$\theta$} &
\multicolumn{1}{c}{$\epsilon$} \\
\hline
  120.0 & 18.46 & 0.32 \\
  140.0 & 20.59 & 0.19 \\
  245.0 & 28.83 & 0.36 \\
  301.0 & 32.83 & 0.44 \\
  351.0 & 36.02 & 0.45 \\
\hline
\end{tabular}
\end{table}

We have, however, found that in the north-west quadrant, i.e. for $x > 0$ and 
$y > 0$, the echo rim is well defined and its shape can be quite well 
reproduced by a circle centred on the star. The results of a least square
fit of a circle centred on the star to the observed echo rim in the
north-west quadrant are given in Table~\ref{echo_nw}. As can be seen from
Table~\ref{echo_nw} the fit is here significantly better (lower $\epsilon$)
than that in Table~\ref{echo_obs}.

\begin{figure}
  \resizebox{\hsize}{!}{\includegraphics{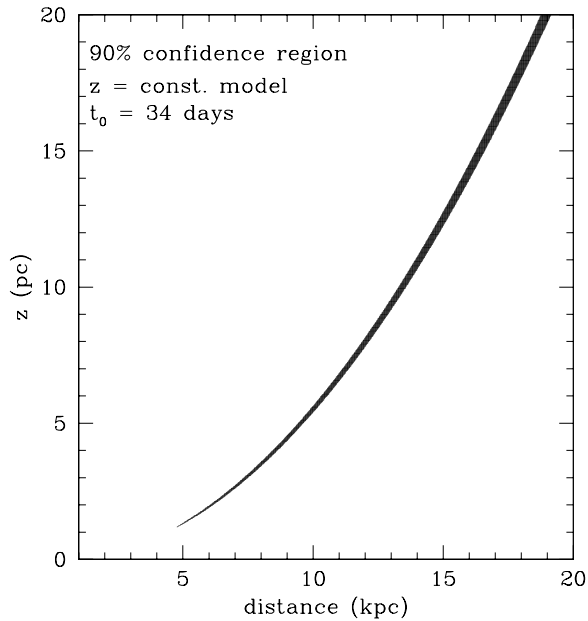}}
  \caption{The 90\% confidence region of the $\chi^2$ test of fitting
            Eq.~(\ref{teta_echo_slab}) to the data
            in Table~\ref{echo_nw}.}
  \label{dz_fit}
\end{figure}

A $\chi^2$ minimum fit of Eq.~(\ref{teta_echo_slab}) to the data
in Table~\ref{echo_nw} gives the best fit for $d \simeq 9.0$~kpc
and $z_0 \simeq 4.5$~pc. The 90\% confidence region from this fit is shown
in Fig~\ref{dz_fit}. Although the 90\% confidence region is more confined
than in the previous case but still the uncertainty in the distance is
large. All what can be said is that the distance to \object{V838~Mon} is greater 
than $\sim 4.8$~kpc.

The echo in the shape of a circle centred on the source can also be modelled
by a spherically symmetric distribution of dust. Therefore we have also
attempted to fit Eq.~(\ref{teta_echo_shell}) to the data in
Table~\ref{echo_nw}. The $\chi^2$ test has however been unable to find a
minimum. It gives acceptable fits starting from $d \simeq 7$~kpc but
finds better and better fits for the distance increasing and increasing, well beyond
reasonable limits. Clearly the test "prefers" the situation in which a
sphere section illuminated by the echo is closer and closer to a plane. This
suggests that the constant $z$ model better approximates the outer edge of
dust in the north--west quadrant of the \object{V838~Mon} echo than the spherical
model.

\section{Structure of the dust region in front of \object{V838~Mon}  \label{dust}}

The structure of the light echo observed in \object{V838~Mon} is rather complex
especially in the last three images. This obviously reflects the complex
distribution of the circumstellar dust. We have attempted to study 
this distribution from an analysis of the available images. Of course only
dust illuminated by the light echo can be studied which practically limits
the analysis only to the regions lying in front of \object{V838~Mon}.

In the available echo images, apart from the outer edge, one can also
distinguish edges of the emission inside the echo. These edges are usually
in colour either blue or red. From the light curve of \object{V838~Mon} it is clear
that the object was bluest in the beginning of the outburst (beginning of
February 2002) while during the fading (mid-April 2002) it was extremely
red (Munari et~al. \cite{munari}, Bond et~al. \cite{bond}). 
Suppose that the paraboloids $ct = 2$ and $ct = 1$ in Fig.~\ref{parab_f}
correspond to the blue beginnng and the red end of the light flash,
respectively. Suppose also that, say, between $z = 5$ and $z = 10$ there is
a dust layer. Then the observer would see an echo ring whose outer edge
would be blue and would correspond to the intersection of $z = 10$ with 
$ct = 2$, while the inner edge would be red and would correspond to 
the intersection of $z = 5$ with $ct = 1$. Thus measurements of the positions 
of the blue and red edges in the light echo of \object{V838~Mon} can be used to estimate
positions of the boundaries of dust layers producing bright regions in the
echo.

We have measured, apart from the outer edge discussed in
Sect.~\ref{dist}, positions of blue and red edges inside the echo
images. This concerns primarily the images taken on Sept.~2, Oct.~28 and
Dec.~17 showing several well defined details in the echo structure. We have
measured only the most obvious features which can be easily identyfied in
all the three images. On the
image from May~20 only the inner red edge has been measured (apart from the
outer blue edge). On Apr.~30 only a B image has been taken so no red edge
can be identified. Following the above considerations we adopt that the blue
edges are produced by the beginning of the outburst so, as discussed in
Sect.~\ref{dist}, they correspond to the light paraboloid with $t_0 = 34$~days.
From the I curve in Bond et~al. (\cite{bond}) it can be found that near
Apr.~17 \object{V838~Mon} declined by factor 2 from its last peak. Therefore we adopt
that the red edges in the light echo correspond to the paraboloid with 
$t_0 = 107$~days.

Knowing $x$ and $y$ from measurements of an edge in 
the echo observed at a given time, $t$,
one can calculate the $z$ coordinate of the dust edge from
Eq.~(\ref{parab_e}). Obviously we measure only the angular values of $x$ and
$y$ so in order to have their absolute values we have to adopt the distance.
Unfortunately, as discussed in the previous section, we have only been able 
to put the lower limit to the distance at $\sim 5$~kpc. 
Munari et~al.(\cite{mundes}) have estimated $d \simeq 10$~kpc from the magnitude 
of the B3V companion to \object{V838~Mon}. However, uncertainty of this estimate is probably
significant. One of the reasons are uncertainties in calibrations of the
absolute magnitudes of the BV stars (e.g. Wegner \cite{wegner}). For the purpose of 
this section we have adopted that \object{V838~Mon} is at a distance of 8~kpc.

\begin{figure*}
\centering
  \includegraphics[width=5.9cm]{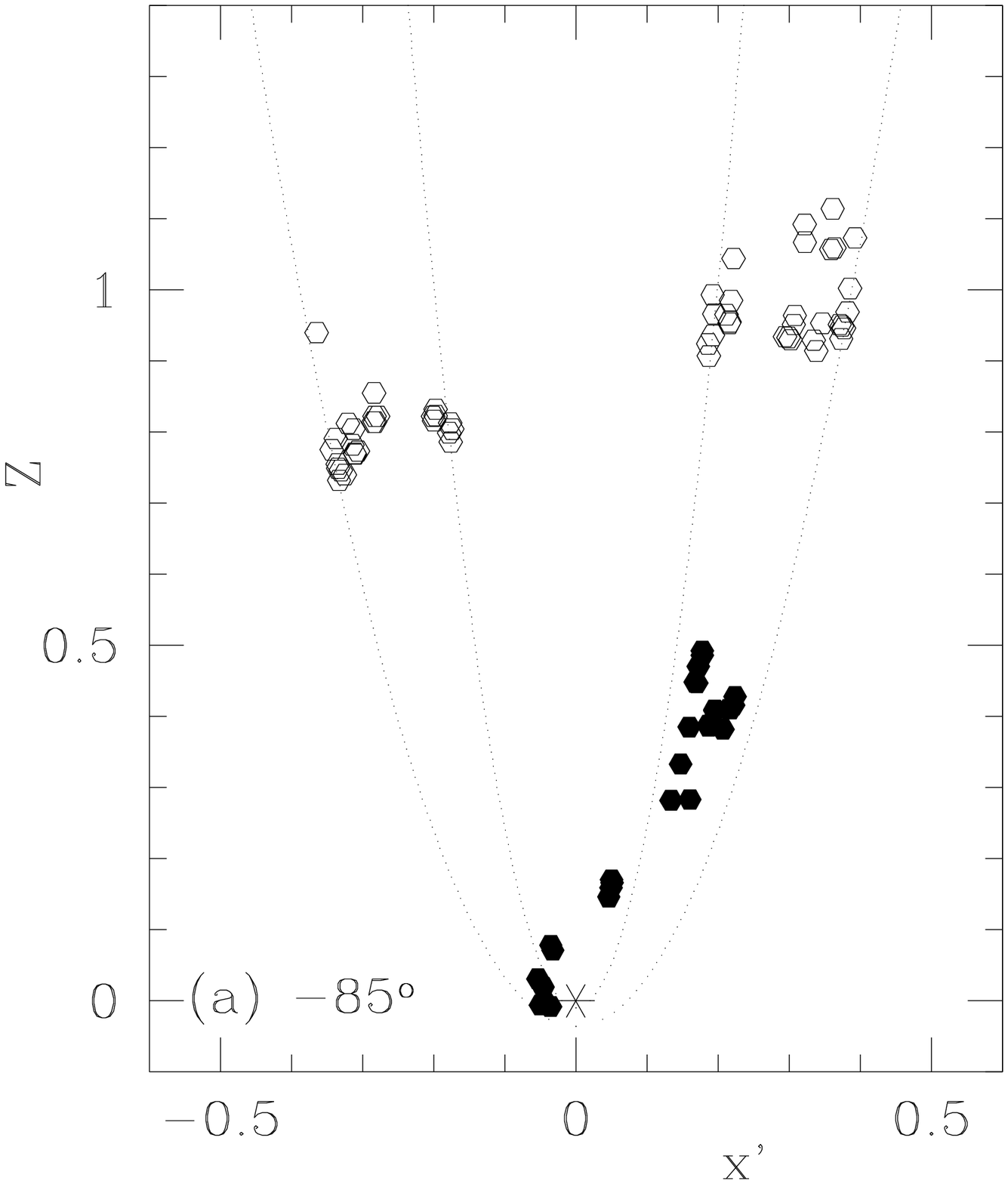}
  \includegraphics[width=5.9cm]{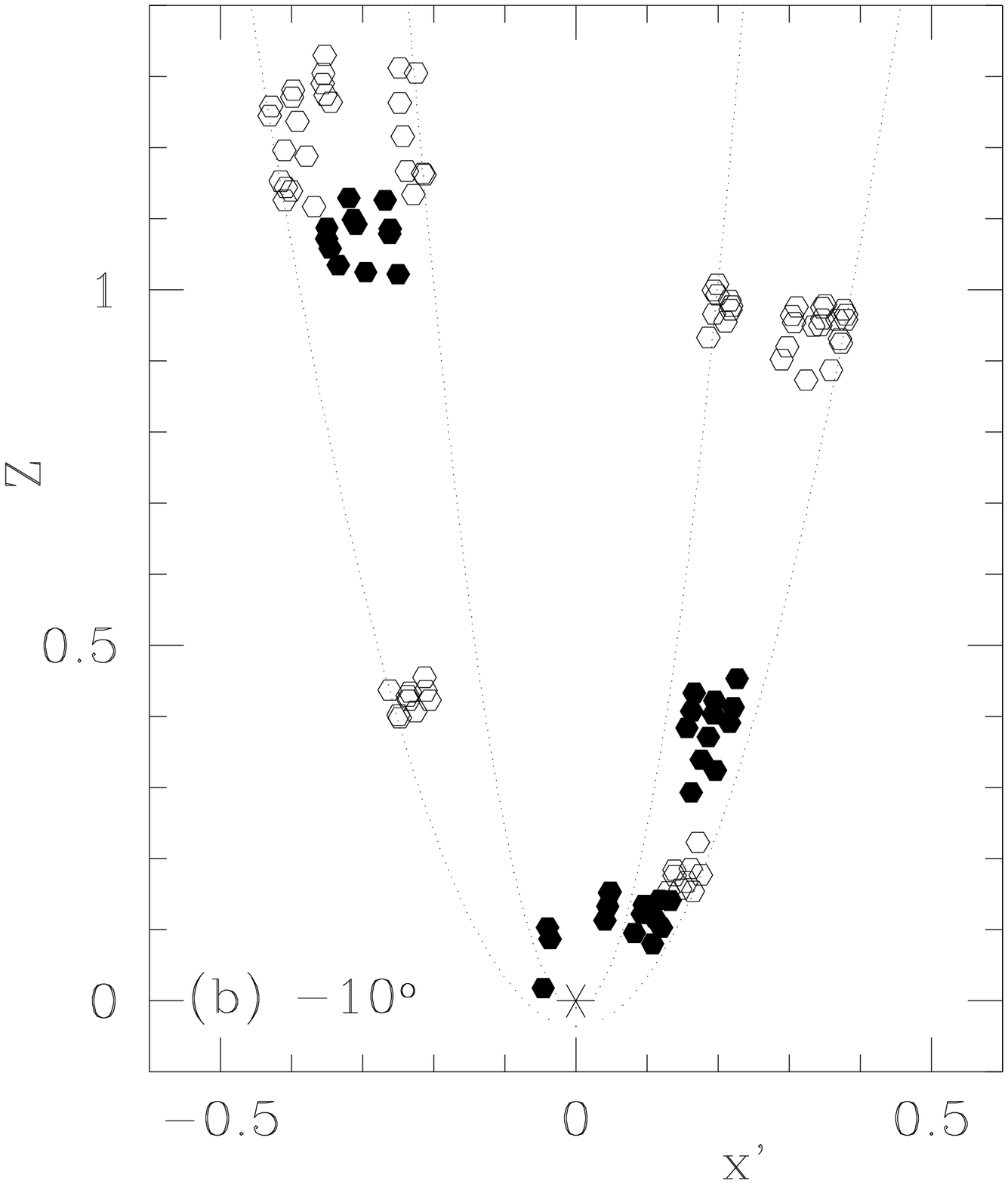}
  \includegraphics[width=5.9cm]{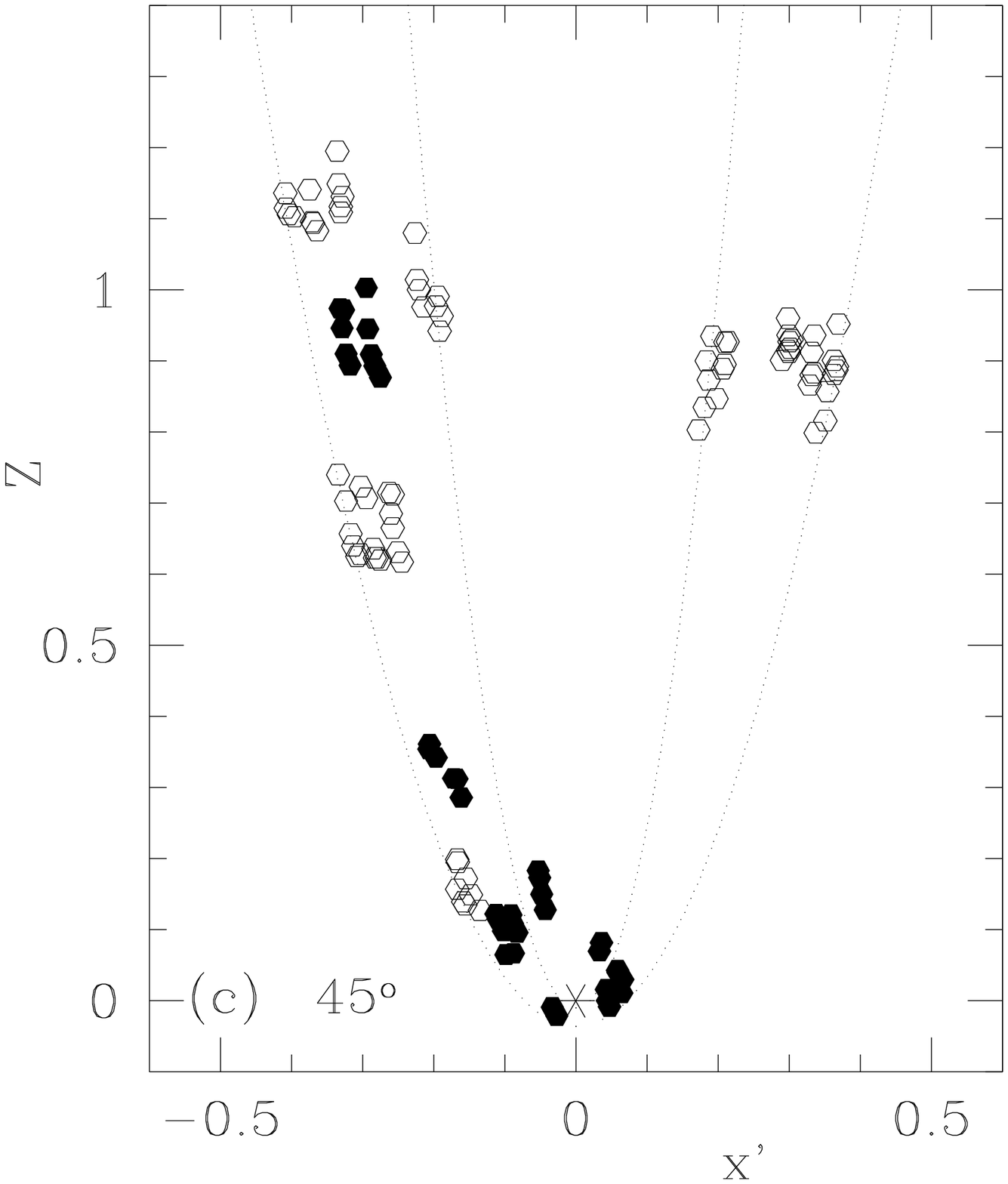}
  \caption{Structure of the dust distribution illuminated by the lihgt echo
of \object{V838~Mon}. In all the images the
ordinate is the $z$ axis. The abscissa, $x'$, is inclined to the $x$ 
(east-west) axis at an angle, $\beta$, (clockwise in respect to the $x$ axis) 
given in the bottom-left corner of each image. Only the points having the
position angle within $\beta \pm 15\degr$ are shown in each image. Open and full 
symbols correspond to upper and lower (in the $z$ coordinate) boundaries of
the dusty regions, respectively. Thin curves show the echo paraboloids
(with $t_0 = 34$~days) on Apr.~30 and Dec.~17.
The axes are in unites of the mean $z$ distance of all the points
in the outer edge of the light echo, which is 3.67~pc. 
The position of the central star is shown by an asterisk.}
  \label{dust_str}
\end{figure*}

The results on the dust structure are
presented in Fig.~\ref{dust_str}. It shows three cuts across the three
dimensional distribution of the dust boundaries inferred from the measurements of
the blue (open points) and red (full points) edges in the light echo. The
plane of each part of Fig.~\ref{dust_str} goes through the $z$ axis and is
inclined to the $x$ (east-west) axis at an angle, $\beta$, (counted clockwise
from the $x$ axis) given in the bottom-left corner of each part of the
figure. Only the points having the
position angle within $\beta \pm 15\degr$ are shown. 
The axes are in unites of the mean $z$ distance of all the points
corresponding to the outer edge of the light echo. This mean $z$ distance is 3.67~pc.
Thin curves show the echo blue paraboloids
(with $t_0 = 34$~days) on Apr.~30 and Dec.~17. Thus, in principle, we can study 
only the dust distribution between these two paraboloids. Exception is the
inner red edge on May~20 whose (red) paraboloid lies within the (blue)
paraboloid on Apr.~30. Dust is present above full and below open symbols (it
is absent above open and below full points).

The cuts shown in Fig.~\ref{dust_str} have been chosen so as to show the
global distribution of dust. They are more or less equally spaced in the
azimuthal angle $\beta$ but they also avoid too detailed structures which
would make discussion too complicated and less clear. The author recommends
consulting the original echo images 
(e.g. at http://hubblesite.org/newscenter/archive/2003/10/) for an easier
understanding of the discussion below.

Fig.~\ref{dust_str}a shows the cut done close to ($5\degr$ inclined to the west
from) the south-north direction. It goes more or less across the major axis
of the central hole in the echo images. The dust distribution in this cut is
very simple. In the southern part ($x' < 0$ in Fig.~\ref{dust_str}a) dust
extends from the hole boundary which lies very close to ($\sim$0.2~pc from) 
the central object up to $\sim$3~pc in $z$. In the northern part ($x' > 0$)
the empty region extends up to $\sim$1.5~pc. Then the dust region goes up to
$\sim$3.5~pc.

The section in Fig.~\ref{dust_str}b goes close to ($10\degr$ inclined to the north
from) the east-west direction. In the eastern part ($x' < 0$) the central
hole is rather compact (similarily as for $x' < 0$ in Fig.~\ref{dust_str}a).
Dust however extends only up to $\sim$1.5~pc. Then above there is an empty
region. Only at $\sim$4~pc there is a thin dust layer. On the western side
of the section ($x' > 0$) the central hole extends up to $\sim$1.5~pc but more
or less in the middle it is cut by a thin layer (or filament) of dust. Then
a thick dust layer extends between $\sim$1.5 and $\sim$3.5~pc.
 
The last section presented in Fig.~\ref{dust_str}c have been done along the
north-east -- south-west direction. In the north-eastern part ($x' < 0$) the
dust structure is rather complex. A rather empty region extends from the
central star up to $\sim$1.0~pc, although roughly in the middle there is a
dusty region. The main dust region extends between $\sim$1.0 and
$\sim$2.5~pc. Note however that this region has its own complex structure as
can be best inferred from the echo image taken on Dec.~17. Finally at a
distance of $\sim$3.7~pc there is a thin dust layer. In the south-western
part an extended dust regions begins at $\sim$0.2~pc from the star and goes
up to $\sim$3.3~pc.

The general picture that emerges from our analysis is as follows. Near the
central object there is a strongly asymmetric region free of dust. Its
boundary is relatively close to the star, i.e. at $\sim$0.2~pc, in the
south-east, south and south-west directions. In the northern direction this
hole extends up to $\sim$1.5~pc. In the north-east and north-west direction
the situation is less clear. The empty region certainly extends to
$\sim$0.5~pc. Futher away there are some dusty regions but rather thin so
one may argue that the central hole extends up to $\sim$1.0--1.5~pc. Above
the central hole there is the main dust region whose outer boundary in the
north, west and south directions lies at $\sim$3.0--3.5~pc. In all the eastern
directions the boundary is closer to the star and near the east axis it is as
close as $\sim$1.5~pc. Finally in the eastern directions there is a thin dust
layer at a distance of $\sim$4~pc.

It should be finally reminded that all the absolute positions of 
the dust regions given
above have been obtained adopting the 8~kpc distance to \object{V838~Mon}. The most
sensitive to the value of distance is the $z$ coordinate which depends
on the square of the distance. Fortunatelly the relative picture of the dust
distribution presented in Fig.~\ref{dust_str} would be hardly affected if
other values of the distance have been adopted.

In Sect. \ref{dist} while constraining the distance to \object{V838~Mon} we have
approximated the outer rim of the observed echo in the north-west quadrant
by a circle centered on the central star and interpreted the results as
produced by a flat surface perpendicular to the line of sight. From the
results in the present section it comes out that it was a reasonable
assumption. The $z$ values of the outer rim points from this quadrant 
(i.e. having $x > 0$, $y > 0$) are scattered around its mean value by less
than 3\% (standard deviation).

\section{Conclusions and discussion}

For all astrophysical objects it is important to know their distances. In
the case of \object{V838~Mon} it is particularly important. 
The outburst of \object{V838~Mon}
does not fit to any known class of outbursting stars so determination of
its luminosity is crucial for attempts to identify the mechanism of the
event. The early estimates of 0.6--0.8~kpc based on a naive interpretation
of the light echo expansion (Munari et~al. \cite{munari}, 
Kimeswenger et~al. \cite{kimes}) led to the maximum luminosity of 
$\sim 10^4 \mathrm{L}_\odot$ which is typical for nova or nova-like
outbursts. Our analysis presented in Sect.~\ref{dist} clearly shows that the
distance and thus also the luminosity are much greater. Unfortunately the
presently available data on the echo evolution do not allow to determine the
distance with a satisfactory precision. But even our lower limit of
$\sim$5~kpc makes the maximum luminosity $\ga 10^6 \mathrm{L}_\odot$
which puts the outburst of \object{V838~Mon} to the most luminous events in our
Galaxy. Unfortunately it seems that we will have to wait long time in order
to obtain a reasonable estimate of the distance from the echo expansion. As
discussed in Sect.~\ref{dist} the condition for this is $ct > z_0$. 
If $z_0 \simeq 3$~pc $t$ should be at least 10~years.

As has been shown in Sect. \ref{dust} the evolution of the light echo can be
used as a very useful tool for studying the dust distribution near the light
source. Although the absolute characteristics of the obtained distribution 
depend on the distance, which is uncertain in the case of \object{V838~Mon}, 
the relative structure is fairly insensitive to this parameter.
What we have done it is a rather simple analysis. A much more detailed study
could be done on absolutely calibrated images of the echo which were not
available to the author. Certainly futher observations would be very
valuable. When the light paraboloid becomes larger and larger, newer and
newer regions of dust are illuminated and more and more complete image of
the dust distribution in the vicinity of the star can be obtained. As can be
seen from Fig.~\ref{dust_str} only a small part of the volume around
\object{V838~Mon} could have been studied so far. But even from it interesting
conclusions can be made. 

We have found no signs of spherical symmetry in
the dust distribution which would have been expected if the observed dust had
resulted from mass loss activities of \object{V838~Mon} in the past. 
The main dust regions has the
outer boundary which is relatively flat and almost perpendicular to the line
of sight in the western part. But near the south-north line it bends toward
the central star or rather splits into a thin outer layer and a
thicker zone closer to the central star. Near the central object there is a
dust free region. This central hole is strongly asymmetric. Its boundary is
quite close to the star in the southern directions but in the northern
directions it is at least 10 times further away (the boundary of the hole in
north seems to be still outside the light paraboloid). The holes near central stars
of e.g. planetary nebulae or HII regions, are often observed. They are
produced by fast winds from the stars which sweep out the
nebular matter. \object{V838~Mon} has a B3V companion (Munari et~al. \cite{mundes}).
It is also plausible that \object{V838~Mon} itself was also of similar spectral type
before the outburst (Tylenda, in preparation). 
According to the standard calibration (Drilling \&
Landolt \cite{drill}) B3V stars have $\log L/L_\odot \simeq 3.5$ which,
using the mean relation of Howarth \& Prinja (\cite{how}), gives 
a mass loss rate of $10^{-9} - 10^{-10} M_\odot/$yr.
Thus it is quite probable that the fast wind from the \object{V838~Mon} system has
created the central hole in the dusty medium. The fact that
the hole is strongly asymmetric implies that \object{V838~Mon} is moving
relatively to the dusty medium. In the image taken 
on Sept.~2 the edge of the central hole in the east, south-east and south
directions is sharp and very bright (compared to the emission behind the
edge). Similar (although less evident) effect is also seen on other images
(May~20, Oct.~28, Dec~17). Most probably we there see the regions compressed
by the wind ahead of the moving star. On Oct.~28 nad Dec.~17 the same edge is
less defined as there is a clear emission between the edge and the central
star. This emission was absent on the images from May~20 and Sept.~2.
Thus on Oct.~28 we have probably started seeing dust
behind the central object. According to Eq.~(\ref{parab_e})
this dust is $\sim$0.1~pc behind the star.

The above finding that V838~Mon is moving relatively to the dusty medium 
is difficult to understand if the
medium were produced by mass loss from \object{V838~Mon} in the past. It is
however quite natural if dust is of interstellar origin. 
We therefore conclude that dust illuminated 
by the light echo of \object{V838~Mon} is most probably of interstellar
origin. This conclusion is also supported by the general lack of spherical
symmetries in the dust distribution, as discussed above. We do not confirm the
assumption made in Bond et~al. (\cite{bond}) that the \object{V838~Mon} light echo
has been produced by "a series of nested spherical dust shells centered on
the star". Consequently their criticism of the merger scenario proposed by
Soker \& Tylenda (\cite{soktyl}) is not relevant. Note, however, that in
view of the distance to \object{V838~Mon} being much larger than 1~kpc the
scenario of Soker \& Tylenda should be revised (which will be done in a
separate paper). Finally, as it is evident from the analysis and discussion
made in the present paper, future
observations of the \object{V838~Mon} light echo are of particular
importance and interest.

\begin{acknowledgements}
The author is very grateful to Noam Soker for his comments on the initial
version of this paper.
\end{acknowledgements}


\begin{thebibliography}{}

   \bibitem[2003]{bond}
Bond, H. E., Henden, A., Levay, Z. G., et~al. 2003, \nat, 422, 405

   \bibitem[2002]{brown}
Brown, N. J. 2002, IAU Circ., 7785

   \bibitem[1986]{cheval}
Chevalier, R. A. 1986, \apj, 308, 225

   \bibitem[1939]{couderc}
Couderc, P. 1939, Ann. d'Astrophys. 2, 271

   \bibitem[2000]{drill}
Drilling, J.S., Landoldt, A. U., 2000, in Allen's Astrophysical Quantities,
the 4th edition, ed. A. N. Cox (Springer-Verlag New York) p. 381

   \bibitem[2002]{henden}
Henden, A., Munari, U., Schwartz, M. B. 2002 IAU Circ., 7859

   \bibitem[1989]{how}
Howarth, I. D., Prinja, R. K. 1989, \apjs, 69, 527

   \bibitem[2002]{kimes}
Kimeswenger, S., Lederle, C., Schmeja, S., Armsdorfer, B. 2002,
\mnras, 336, L43

   \bibitem[2002b]{mundes}
Munari, U., Desidera, S., Henden, A. 2002b, IAU Circ., 8005

   \bibitem[2002a]{munari}
Munari, U., Henden, A., Kiyota, S., et~al. 2002a, \aap, 389, L51

   \bibitem[2003]{retmar}
Retter, A., Marom, A. 2003, \mnras, submitted

   \bibitem[2003]{soktyl}
Soker, N., Tylenda, R. 2003, \apjl, 582, L105

   \bibitem[1994]{sparks}
Sparks, W. 1994, \apj, 433, 19

   \bibitem[2000]{wegner}
Wegner, W. 2000, \mnras, 319, 771

   \bibitem[1995]{xu}
Xu, J., Crotts, A. P. S., Kunkel, W. E. 1995, \apj, 451, 806

\end{thebibliography}
\end{document}